\documentclass[english,aps, reprint, prb, amsmath,amssymb,floatfix,superscriptaddress,eqsecnum,longbibliography]{revtex4-2}

\newcommand{\bra}[1]{\langle{#1}|}
\newcommand{\ket}[1]{|{#1}\rangle}

\usepackage[utf8]{inputenc}
\usepackage{booktabs}
\usepackage{multirow}
\usepackage[english]{babel}
\usepackage{amssymb,amsmath,mathrsfs}
\usepackage{bm}
\usepackage{graphicx}
\usepackage{ulem}
\usepackage[pdftex,colorlinks=true]{hyperref}
\usepackage{braket}
\usepackage{color}
\usepackage{tabularx}
\usepackage{units}


\begin{document}
\title{Coupled spin-lattice dynamics from the tight-binding electronic structure}
\author{Ramon Cardias}
\affiliation{Department of Applied Physics, School of Engineering Sciences, KTH Royal Institute of Technology, AlbaNova University Center, SE-10691 Stockholm, Sweden}
\affiliation{Instituto de Física, Universidade Federal Fluminense, 24210-346, Niterói RJ, Brazil}
\author{Simon Streib}
\affiliation{Department of Physics and Astronomy, Uppsala University, Box 516,
SE-75120 Uppsala, Sweden}
\author{Zhiwei Lu}
\affiliation{Department of Applied Physics, School of Engineering Sciences, KTH Royal Institute of Technology, AlbaNova University Center, SE-10691 Stockholm, Sweden}
\author{Manuel Pereiro }
\affiliation{Department of Physics and Astronomy, Uppsala University, Box 516,
SE-75120 Uppsala, Sweden}
\author{Anders Bergman}
\affiliation{Department of Physics and Astronomy, Uppsala University, Box 516,
SE-75120 Uppsala, Sweden}
\author{Erik Sj\"oqvist }
\affiliation{Department of Physics and Astronomy, Uppsala University, Box 516,
SE-75120 Uppsala, Sweden}
\author{Cyrille Barreteau}
\affiliation{Universit\'e Paris-Saclay, CEA, CNRS, SPEC, 91191, Gif-sur-Yvette, France}
\author{Anna Delin }
\affiliation{Department of Applied Physics, School of Engineering Sciences, KTH Royal Institute of Technology, AlbaNova University Center, SE-10691 Stockholm, Sweden}
\affiliation{ Swedish e-Science Research Center (SeRC), KTH Royal Institute of Technology, SE-10044 Stockholm, Sweden}
\affiliation{Wallenberg Initiative Materials Science for Sustainability (WISE), KTH Royal Institute of Technology, SE-10044 Stockholm, Sweden}
\author{Olle Eriksson }
\affiliation{Department of Physics and Astronomy, Uppsala University, Box 516,
SE-75120 Uppsala, Sweden}
\affiliation{Wallenberg Initiative Materials Science for Sustainability (WISE), Uppsala University, Box 516,
SE-75120 Uppsala, Sweden}

\author{Danny Thonig}
\affiliation{School of Science and Technology, \"Orebro University, SE-70182 Örebro,
Sweden}
\affiliation{Department of Physics and Astronomy, Uppsala University, Box 516,
SE-75120 Uppsala, Sweden}
\date{\today}

\begin{abstract}
We developed a method which performs the coupled adiabatic spin and lattice dynamics based on the tight-binding electronic structure model, where the intrinsic magnetic field and ionic forces are calculated from the converged self-consistent electronic structure at every time step. By doing so, this method allows us to explore limits where the physics described by a parameterized spin-lattice Hamiltonian is no longer accurate. We demonstrate how the lattice dynamics is strongly influenced by the underlying magnetic configuration, where disorder is able to induce significant lattice distortions. 
The presented method requires significantly less computational resources than \textit{ab initio} methods, such as time-dependent density functional theory (TD-DFT). Compared to parameterized Hamiltonian-based methods, it also describes more accurately the dynamics of the coupled spin and lattice degrees of freedom, which becomes important outside of the regime of small lattice and spin fluctuations.
\end{abstract}
\maketitle

\section{Introduction}
The interplay between spin and lattice degrees of freedom holds paramount significance in various fields of condensed matter physics, notably in ultrafast dynamics \cite{beaurepaire_ultrafast_1996,stockem_anomalous_2018,kormann_temperature_2014,pradip_lattice_2016,perera_collective_2017}. At extremely low temperatures, ionic motion has been demonstrated to be negligible compared to spin motion~ \cite{antal_first-principles_2008,niu_adiabatic_1999,qian_spin_2002,halilov_adiabatic_1998}. Hence, in this regime, the spin and lattice degrees of freedom are treated independently, which effectively describes a wide range of applications \cite{eriksson_atomistic_2017}. However, a growing number of phenomena of interest to the scientific community necessitate consideration of magnon-phonon coupling when magnon and phonon frequencies are of similar magnitude: \textit{i)} In multiferroics, the polarization induced by the combination of the charge and/or magnetic non-collinearity distorts the lattice, making the spin-lattice coupling the pivotal mechanism behind the magnetoelectric effect~\cite{pradip_lattice_2016,sushkov_electromagnons_2008,pimenov_possible_2006,kimura_distorted_2003}; \textit{ii)} In spintronics applications and for terahertz applications, angular momentum currents can be converted between lattice and spin degrees of freedom via the spin-lattice coupling~\cite{streib_damping_2018,an_coherent_2020,ruckriegel_long-range_2020}; \textit{iii)} Moreover, in femto-second ultrafast demagnetization processes it is debated that the angular momentum transfer is mediated by a coupling between lattice, spin, and electronic degrees of freedoms~\cite{tauchert_polarized_2022,dornes_ultrafast_2019}.

In the adiabatic approximation, both forms of dynamics are well understood~\cite{eriksson_atomistic_2017,leimkuhler_molecular_2015}. The magnetization dynamics usually takes the form of the Landau-Lifshitz-Gilbert (LLG) equation~\cite{landau_theory_1935,gilbert_phenomenological_2004}. In such simulations, one often introduces 
a coupling to a thermal reservoir, through the Langevin dynamics (LD) approach, which ensures that the distribution of the energy of atomic spins follows a Boltzman distribution\cite{Skubic_2008,eriksson_atomistic_2017}. Regarding the lattice dynamics, it follows the principles of Newtonian dynamics~\cite{leimkuhler_molecular_2015}. In this case, the time evolution of the ionic positions is driven by the atomic forces at every time step. Different techniques to minimize the atomic forces  can be found in the literature~\cite{gronbech-jensen_simple_2013}, in order to lead the system to its respective relaxed structure. Similar techniques are also applied to minimize torques in the magnetic system \cite{BESSARAB2015335}.

In recent works, progress has been made towards coupling of these two dynamical processes. One way is to consider that the spin-spin exchange parameters, e.g. the isotropic ($J_{ij}$) and Dzyaloshinskii-Moriya ($\vec{D}_{ij}$) interaction, depend on the atomic displacements, where $i,j$ are the sites indices. This approach has been considered and applied in Ref.~\cite{hellsvik_general_2019}, where a Taylor expansion of an effective spin Hamiltonian was considered in order to collect these contributions, always assuming the adiabatic limit. Such expansion breaks the rotational invariance and conservation of angular momentum~\cite{garanin_conservation_2021,PhysRevB.108.L060404}, which however may be restored by expressing the expansion using instantaneous differences between atomic positions instead of displacements~\cite{PhysRevB.108.L060404,Sadhukhan2023}. However, the use of displacements is a practical way to compute the parameters of various terms in the spin-lattice Hamiltonian, e.g. the force constants, and, therefore, this approach will be used in the present work. 

Moreover, in Ref.~\cite{stockem_anomalous_2018} classical spin dynamics and \textit{ab initio} molecular dynamics were coupled in order to study CrN, where the authors were able to identify non-adiabatic effects, beyond an effective Hamiltonian description, from both the dynamics of the lattice as well as from the spin. When using such an effective Hamiltonian to describe the time evolution of the magnetic moments, it is assumed that their motion do not affect the parameters calculated from the initial state. Recent literature, however, has shown that this assumption only holds for small fluctuations around the ground state~\cite{cardias_bethe-slater_2017,cardias_spin_2021,cardias_first-principles_2020,cardias_dzyaloshinskii-moriya_2020,streib_exchange_2021,streib_exchange_2021,szilva_interatomic_2013,szilva_quantitative_2022,kvashnin_microscopic_2016,szilva_theory_2017,jacobsson_efficient_2022}. It can be understood that in strongly non-collinear magnetic states, the electrons cause a spin current by hop between the atomic sites of the sample, giving rise to high-order magnetic interactions (or multi-spin parameters)~\cite{brinker_generalization_2022,brinker_prospecting_2020,dos_santos_dias_proper_2021}. Another way to look at this phenomenon is that the interactions between two sites only depend on their current magnetic configuration~\cite{cardias_first-principles_2020,cardias_dzyaloshinskii-moriya_2020,szilva_interatomic_2013}. Therefore, it is expected that for systems where it becomes highly relevant to treat strong non-collinear magnetic configurations, the spin-Hamiltonian assumption might break or miss relevant features in magnon and phonon spectra compared to experiment~\cite{szilva_interatomic_2013,rodrigues_finite-temperature_2016}. Therefore, as far as computational tools go to describe spin-lattice dynamics of a given system, Hamiltonian based models rely on the correct description of the system electronic structure throughout the parameters, which can be a delicate job for materials with non-collinear magnetism, while \textit{ab initio} methods can only treat a limited number of atoms, due to the heavy computational cost.

This paper presents a solution that involves computing the key components of the interconnected spin-lattice dynamics, including molecular forces and effective magnetic fields, directly from the electronic structure. The approach utilized here is based on a Slater-Koster parameterized tight-binding model using the National Research Lab (NRL) formalism~\cite{slater_simplified_1954,mehl_applications_1996,barreteau_efficient_2016}, to calculate the ground state of a given system at every time step, from which we evaluate the molecular forces via the Hellmann-Feynman theorem~\cite{feynman_forces_1939,hellmann_einfuhrung_1937,popov_extension_1998,cabrera_equivalente_1990} and the effective magnetic field as the field opposing the constraining magnetic field \cite{stocks_towards_1998,streib_equation_2020}. 
Based on this, we performed coupled spin-lattice dynamics of Cr and Fe clusters (triangular and nanochains with 10 atoms) to find the ground state, when varying both but also only one degree of freedom. From this ground state, we explicitly calculate the parameter for an effective Hamiltonian and analyze how these parameters -- force constants and exchange coupling parameters -- vary in different situations and magnetic configurations. Our results show that for small variations near the magnetic ground state, these parameters are fairly constant. Far away from the ground state, particularly for the Cr triangle case where the magnetic ground state is non-collinear, there is a significant difference. As to the magnetic properties, Cr tends to stabilize in a AFM fashion, which results in the Néel AFM magnetic structure for the triangular trimer and a collinear AFM for the nanochains. The Fe based systems present a FM ordering for the triangular trimer and a spin-spiral like configuration for the nanochains, resulted from the interplay between short and long range interactions. During the relaxation process of the atomic positions, we show how the magnetic moment length varies and how the lattice dynamics is affected by the motion of the magnetic moments. In fact, for the Cr trimer, our results reveal that a distorted magnetic structure can lead to a distorted lattice structure. Lastly, we compared our tight-binding spin-lattice dynamics with semi-classical spin-lattice dynamics from the effective Hamiltonian, which revealed missing contributions to the effective Hamiltonian in the highly disordered case.

This paper is structured as the following: in Sec.~\ref{sec:asld}, we discuss about the adiabatic spin-lattice dynamics and introduce the equations regarding the time evolution of the magnetic moments and ionic positions cause by constraining fields and forces, directly obtained from the underlying electronic structure. In Sec.~\ref{sec:sldham}, we introduce the semi-classical effective spin-lattice Hamiltonian which is used to compare the tight-binding with the semi-classical spin-lattice dynamics and for which we calculated respective parameter. These parameters are discussed and analyzed for the cluster of Fe and Cr in Sec.~\ref{sec:resultssldham}. Finally, in Sec.\ref{sec:sldres}, the spin-lattice dynamics of these nanoclusters obtained from tight binding and from an effective Hamiltonian are presented. We summarize our findings in Sec.~\ref{sec:sumdis}.


\section{Adiabatic spin-lattice dynamics \label{sec:asld}}

Following our assumption in Ref. \cite{streib_adiabatic_2022}, we are interested to describe physics of magnetic moments $\textbf{e}$ and lattice displacements $\mathbf{u}$ on a timescale above $\unit[1]{fs}$. Here, adiabatic dynamics operates under the premise that electrons evolve significantly faster than the dynamics of the respective degrees of freedom ~\cite{antropov_spin_1996,halilov_adiabatic_1998}. This premise is solidly grounded, particularly for quasi-particle excitations possessing energies considerably below intrinsic electron energies like Stoner spin splitting and variations in electron hopping. 
On comparable energies, the dynamics of the electron need to be addressed from first principle methods, e.g., time-dependent density functional theory (TD-DFT)~\cite{tancogne-dejean_time-dependent_2020,qian_spin_2002}. 

In the adiabatic approximation, the total energy of the system is a function of only the magnetic moment directions $\left\{\mathbf{e}\right\}$ and lattice displacements $\left\{\mathbf{u}\right\}$, where the electronic degrees of freedom can be regarded as being in a quasi-equilibrium state,
\begin{equation}
E =E(\{\mathbf{e}\},\{\mathbf{u}\}).\label{eq:adiabatic energy}
\end{equation}
For precise computation of electronic states and, consequently, the total energy within first-principles methods, it is imperative to rigorously constrain the calculated magnetic moments with specified `input' moment directions denoted as ${\mathbf{e}}$. Without this constraint, the system would naturally experience a finite torque and revert to the absolute ground state due to relaxation processes~\cite{stocks_towards_1998,ujfalussy_constrained_1999,ma_constrained_2015}. Adding an extra term representing the constraining field $\mathbf{B}_{i}^{\text{con}}$ at a lattice site $i$ to the electron Hamiltonian \cite{streib_equation_2020,streib_adiabatic_2022}, we conserve calculated magnetic moments in the direction ${\mathbf{e}}$.   Here, $\mathbf{B}_{i}^{\text{con}}$ is obtained according to $\mathbf{B}_{i}^{\text{con}}\cdot\mathbf{e}_{i}=0$ in an iterative, self-consistent algorithm ~\cite{stocks_towards_1998,ujfalussy_constrained_1999}. This criterion for the constraining field shows that its acts perpendicularly to the magnetic moment direction $\mathbf{e}_{i}$ and, thus, do not change the magnetic moment length. 

In the constrained electronic solution, the magnetic moment experiences an intrinsic effective field $\mathbf{B}_{i}^{\text{eff}}=-\mathbf{B}_{i}^{\text{con}}$ \cite{streib_adiabatic_2022}. This field drives the time evolution of the magnetic moment $\mathbf{m}_i=m_i\mathbf{e}_i$ following the Landau Lifshitz Gilbert equation~\cite{antropov_spin_1996,eriksson_atomistic_2017}:
\begin{align}
\dot{\mathbf{e}}_{i} & =\frac{\gamma}{1+\alpha^2}\mathbf{e}_{i}\times\mathbf{B}_{i}^{\text{eff}}+\frac{\alpha\gamma}{1+\alpha^2}\mathbf{e}_{i}\times\left(\mathbf{e}_{i}\times\mathbf{B}_{i}^{\text{eff}}\right),\label{eq:llg}
\end{align}
where $\gamma$ is the gyromagnetic ratio, $\alpha$ is the phenomenological Gilbert damping. We keep the Gilbert damping finite due to dissipation effects by spin disorder or electron correlation effects \cite{PhysRevB.91.165132}. For instance, the origin of the Gilbert damping has been discussed to originate from spin-orbit coupling~\cite{PhysRevLett.102.137601}. 
 
Regarding the dynamics of the lattice degree of freedom, the Born-Oppenheimer approximation can be considered, where the electronic and ionic degrees of freedoms can be treated separately. Here, forces $\mathbf{F}_k$ acting on the atom $k$ are well defined by the Hellmann-Feynman forces~\cite{feynman_forces_1939,hellmann_einfuhrung_1937,popov_extension_1998,cabrera_equivalente_1990} and the adiabatic lattice dynamics simply consists in solving numerically Newton's equation of motion to produce the trajectory $\textbf{u}_{k}$ of the atom with mass $M_{k}$ and momentum $\textbf{p}_{k}$. Also here, a friction force proportional to the momentum and scaled by the parameter $\nu$~\cite{gronbech-jensen_simple_2013} is added, analogous to the damping constant $\alpha$ in the case of spin-dynamics,

\begin{align}
    \dot{\mathbf{u}}_{k} = & \frac{\mathbf{p}_{k}}{M_{k}}, \label{eq:md0} \\
    \dot{\mathbf{p}_{k}} = & \mathbf{F}_{k} - \nu\mathbf{p}_{k},
    \label{eq:md}
\end{align}
where $\nu$ is the friction force. It should be noted that we ignore the influence of temperature in both equations of motion, that e.g. in Langevin dynamics is introduced via a stochastic field.

In our work we represent the electronic structure by a non-orthogonal Slater-Koster parametrized, realspace tight-binding model implemented in the software package \textit{Cahmd}\cite{CAHMD}. The Slater-Koster parameters are expanded in terms of the hopping distance of the electrons between the atoms according to the NRL tight-binding method. Magnetism is included by a Stoner term proportional to the magnetic moments $\mathbf{e}$. A local charge neutrality term conserves the total charge of the system. Since the orbital overlap matrix enters via Mulliken transformation \cite{Schena2010} also into the magnetic as well as charge neutrality terms of the Hamilonian, not only the hopping but also the exchange splitting and charge distribution in the system become dependent on the lattice displacement. More details about the model are given in the supplemental material \ref{app:tbmodel} and in Ref.~\cite{mehl_applications_1996,barreteau_efficient_2016}.
The electronic structure is self-consistently iterated with respect to the charge, the magnetic moment length, and the constraining field up to the accuracy of $1\cdot 10^{-10}$, i.e. the relative error of each convergence parameter (charge, magnetic moment length and constraining field) is of that order. From the solution we extract the effective field \cite{streib_equation_2020} and calculate from the Hellman-Feynman theorem the lattice forces \cite{hellmann_einfuhrung_1937, jdziedzic2007}. More details in how the effective magnetic field and ionic forces, used in Eqs.~(\ref{eq:llg}) and (\ref{eq:md}), are calculated can be found in Ref.~\cite{streib_equation_2020} and in Appendix.~\ref{sec:forces}, respectively. It should be noted that we neglect terms in the effective fields and forces that are proportional to the variation of the self-consistent parameters, say changes of the charges with the respective degree of freedom  $\nicefrac{\partial n}{\partial \vec{e}}$, $\nicefrac{\partial n}{\partial \vec{u}}$ as well as changes of the magnetic moment length with the respective degree of freedom $\nicefrac{\partial m}{\partial \vec{e}}$, $\nicefrac{\partial m}{\partial \vec{u}}$. These gradients are typically small and negligible. 

Having the effective fields and forces, we integrate Eq.~(\ref{eq:llg}) and Eq.~\eqref{eq:md} using the implicit mid-point method as described in Ref.~\cite{hellsvik_general_2019} using parameters $\alpha=0.1$ and $\nu=\unit[0.1]{fs^{-1}}$. Note that a common way to solve Eqs.~(\ref{eq:md0}-\ref{eq:md}) is to use the so-called Verlet-type based algorithms~\cite{gronbech-jensen_simple_2013}. However, these methods, for the purpose of spin-lattice dynamics, produce numerical instabilities which lead to the non-conservation of the total energy of the system. Our numerical integration step is $\Delta t =\unit[1]{as}$

\section{The spin-lattice Hamiltonian\label{sec:sldham}}

A well defined procedure is to project the energy in Eq.~\eqref{eq:adiabatic energy} to a parametrized Hamiltonian. Following Ref.\cite{hellsvik_general_2019}, such a Hamiltonian can be defined as

\begin{align}
	\mathcal{H_{SLD}}&=-\frac{1}{2}\sum_{ij}\mathcal{J}_{ij}^{\alpha\beta}e_{i}^{\alpha}e_{j}^{\beta}-\frac{1}{2}\sum_{ijk} \label{eq:sldham}{\Gamma_{ijk}^{\alpha\beta\mu}u_{k}^{\mu}e_{i}^{\alpha}e_{j}^{\beta}} \\ \nonumber &+\frac{1}{2}\sum_{kl}{\Phi_{kl}^
	{\mu\nu}}u_{k}^{\mu}u_{l}^{\nu}+\frac{1}{2}\sum_{k}{M_{k}\nu_{k}^{\mu}\nu_{k}^{\mu}}, \\ \nonumber
\end{align}
where $\mathcal{J}_{ij}^{\alpha\beta}$ is the exchange tensor, $\Gamma_{ijk}^{\alpha\beta\mu}$ is the derivative of the exchange tensor with respect to the lattice displacement $u_{k}^{\mu}$, $\Gamma_{ijk}^{\alpha\beta\mu}=\frac{\partial \mathcal{J}_{ij}^{\alpha\beta}}{\partial u_{k}^{\mu}}$, $\Phi_{kl}^{\mu\nu}$ is the force constant and the last term is the kinetic energy. In our study, no spin-orbit coupling is considered such as that the exchange tensor $\mathcal{J}_{ij}^{\alpha\beta}$ has only diagonal, identical terms. It is important to highlight that in the exchange striction constants $\Gamma_{ijk}$ for nanostructure, the term $i=j$ is finite and is required to fulfill Newton's third law for the exchange striction term in the ferromagnetic collinear state: $\sum_{ij}{\Gamma_{ijk}^{\mu}}=0$ for all $k$. From Eq.~\eqref{eq:sldham} and for performing semi-classical spin-lattice dynamics, one can calculate the effective magnetic field $\mathbf{B}_{i}^{\text{eff}}=-\nicefrac{1}{m_i}\nicefrac{\partial \mathcal{H_{SLD}}}{\partial \mathbf{e}_{i}}$ and the ionic force $\mathbf{F}_{k}=-\nicefrac{\partial \mathcal{H_{SLD}}}{\partial \mathbf{u}_{k}}$ by taking the derivative of the spin-lattice Hamiltonian with respect to the magnetic moment of site $i$ and displacement of site $k$, respectively \cite{streib_adiabatic_2022,stockem_anomalous_2018,cardias_spin_2021,eriksson_atomistic_2017}.



In order to compare the dynamics with effective fields and forces from tight binding electronic structure and from semi-classical parametrization, we calculate the parameters in Eq.~\eqref{eq:sldham} also from the tight binding method. In Ref.\cite{streib_exchange_2021}, we solved initial challenges to calculate the spin exchange for an arbitrary non-collinear magnetic configuration. This study is based on the original work by Lichtenstein~\textit{et al}~\cite{katsnelson_first-principles_2000,katsnelson_magnetic_2004}. The exchange striction term is calculated via the numerical procedure described in Ref.~\cite{hellsvik_general_2019}. On the other hand, the exchange striction parameter can also be calculated explicitly via Green's function methods as demonstrated in Ref.~\cite{PhysRevLett.129.067202}. The numerical procedure, though, allows to calculate also the exchange striction for any arbitrary magnetic state due to corrections coming from the constraint methodology. 



The force constants are obtained from the numerical derivation of the forces
\begin{equation}
    \Phi_{kl}^{\mu\nu} = \frac{\partial F_{k}^{\mu}}{\partial u_{l}^{\nu}},
\end{equation}
where $\{\mu,\nu\}$ are the cartesian coordinates and $\{k,l\}$ are the atomic sites. The $\Phi$'s are symmetrized according to point group symmetrical relations. For instance, as a consequence of the Newton's third law, the force constants have the following sum rule

\begin{equation}
    \Phi_{kk}^{\mu\nu} = - \sum_{k\neq l}\Phi_{kl}^{\mu\nu}.
    \label{eq:sumrule}
\end{equation}
Also here, the force constants can be explicitly calculated from the Green's function of the underlying electronic structure as proposed by Fransson \textit{et al}~\cite{fransson_microscopic_2017}.

In Eq.\eqref{eq:sldham}, we explicitly neglect the expansion of the force constants with respect to the magnetic moment configuration, following Ref.~\cite{PhysRevLett.113.165503}. These are terms of the form

\begin{align}
   \Pi_{kli}^{\mu\nu\alpha} & = \frac{\partial \Phi_{kl}^{\mu\nu}}{\partial e_{i}^{\alpha}}e_{i}^{\alpha}u_{k}^{\mu}u_{l}^{\nu}, \label{eq:phi1}\\
    \Omega_{klij}^{\mu\nu\alpha\beta} & = \frac{\partial^{2} \Phi_{kl}^{\mu\nu}}{\partial e_{i}^{\alpha}\partial e_{j}^{\beta}}e_{i}^{\alpha}e_{j}^{\beta}u_{k}^{\mu}u_{l}^{\nu}.\label{eq:phi2}
\end{align}
To nevertheless prove the existence of such high-order terms, we calculate the changes of the force constants when rotating the magnetic moment.

Due to the explicit form of the constrained isotropic spin-spin exchange \cite{streib_exchange_2021}, we are able to also time-resolve $\mathcal{J}_{ij}^{\alpha\beta}$. 
But the numerical treatment of the force constants as well as exchange striction terms, do not allow for a time resolved output during the spin-lattice tight binding dynamics. Calculating the parameters of the spin-lattice Hamiltonian \eqref{eq:sldham} in the ground state of the system, we are also able to perform semi-classical dynamics based on the $\mathcal{H_{SLD}}$. It should be noticed that calculations of the parametrization as well as the semi-classical dynamics are also done via the package \textit{Cahmd} \cite{CAHMD}. Thus, the entire methodology presented is on equal footing.






\section{Parametrization of the spin-lattice Hamiltonian}\label{sec:resultssldham}

In the following we will first discuss the parameterization of the spin-lattice Hamiltonian \eqref{eq:sldham} for trimers and nanochains of Fe and Cr. These parameters are required to calculate the effective fields and forces from the Hamiltonian \eqref{eq:sldham} and to perform semi-classical dynamics. Besides analyzing the parameters, we will also discuss the magnetic moment dependent force constants just for the trimer case, to confirm whether the terms \eqref{eq:phi1} and \eqref{eq:phi2} can be neglected.

\begin{figure}
    \centering
    \includegraphics[width=0.2\columnwidth]{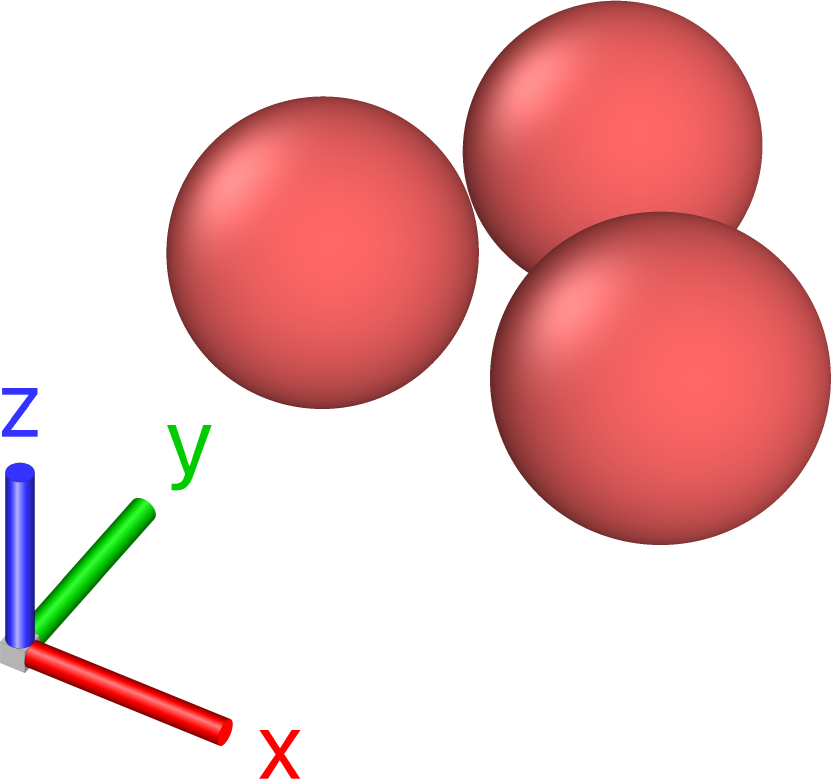}\\
    (a) Trimer\\[10pt] 
    
    \includegraphics[width=0.7\columnwidth]{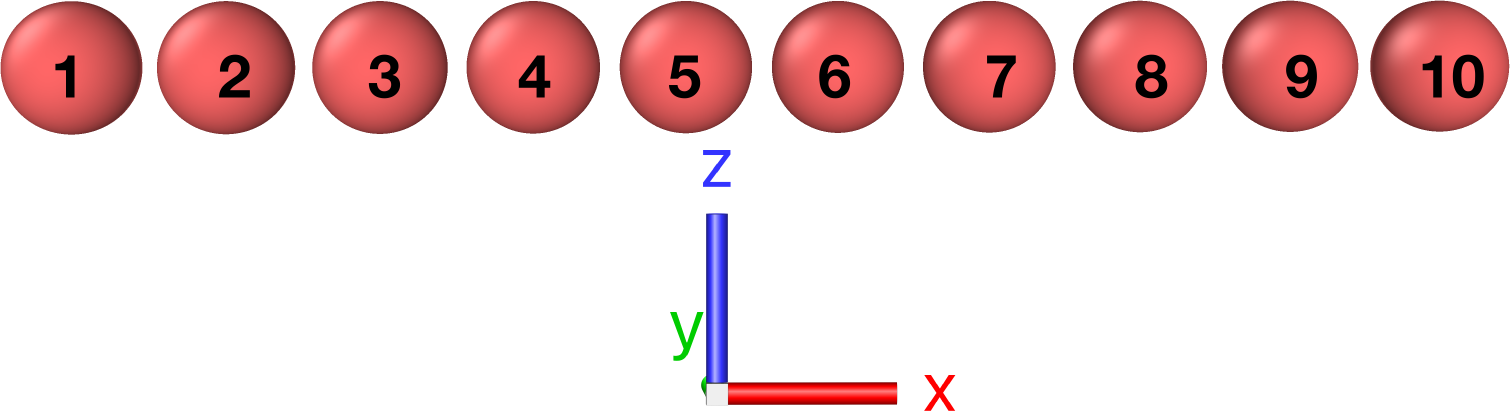}\\
    (b) Nanowire with 10 atoms
    
    \caption{Schematic representation of (a) the triangle trimer and (b) the nanowire with 10 atoms, with their respective coordinate system represented by the tripod.}
    \label{fig:schematic}
\end{figure}

\subsection{Triangular trimer}

Here, we calculate the spin-lattice Hamiltonian parametrization using the magnetic and lattice ground state as reference state.
The ground state was obtained by performing the tight binding spin-lattice dynamics starting from an arbitrary state (see next Section). Due to the large energy dissipation, the system converges rapidly ($\approx \unit[500]{fs}$) to its zero-temperature ground state. In general, the dissipation time of each channel (lattice or spin) will depend on how far each channel is from its respective ground state. For instance, for the Cr trimer case, since the ground state is non-collinear, the magnetic moments takes longer to reach equilibrium compared to the lattice counterpart.
Here the magnetic ground state found is ferromagnetic (FM) and Néel antiferromagnetic (N-AFM) state for Fe and Cr, respectively. In the case of Fe, the isotropic spin-spin exchange is $J_{12}=\unit[362.8]{meV}$, force constants are $\Phi_{12}^{xx}=\unit[-5.92]{eV/a.u.^2}$,$\Phi_{12}^{xy}=  \unit[-0.40]{eV/a.u.^2}$, and $\Phi_{12}^{yy}=\unit[0.23]{eV/a.u.^2}$ in an irreducible representation. For the equilateral triangle sitting in the xy-plane, the elements $\Phi^{xz},\Phi^{zx}$ and $\Phi^{yz},\Phi^{zy}$ are zero by symmetry. The exchange striction terms are calculated (data not shown, but used in Section~\ref{sec:resultssldham}). The magnitude of $\Gamma$ is comparable to the values of the spin-spin exchange or larger as also found in Ref. \cite{hellsvik_general_2019}. Other symmetries mentioned in Ref.\cite{hellsvik_general_2019} are also confirmed.

\begin{figure}
    \centering
    \includegraphics[width=1\columnwidth]{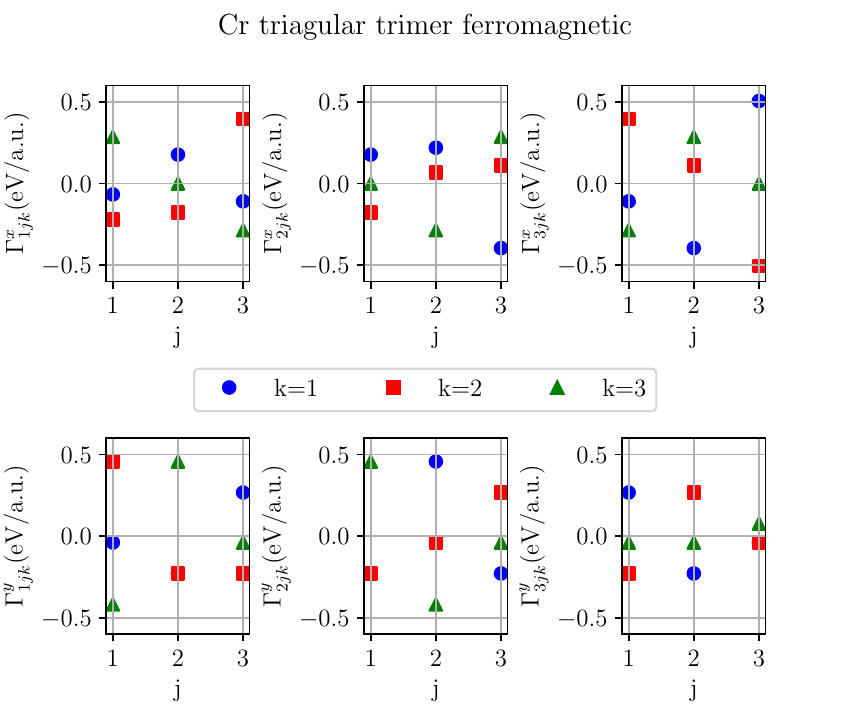}\\
    \includegraphics[width=1\columnwidth]{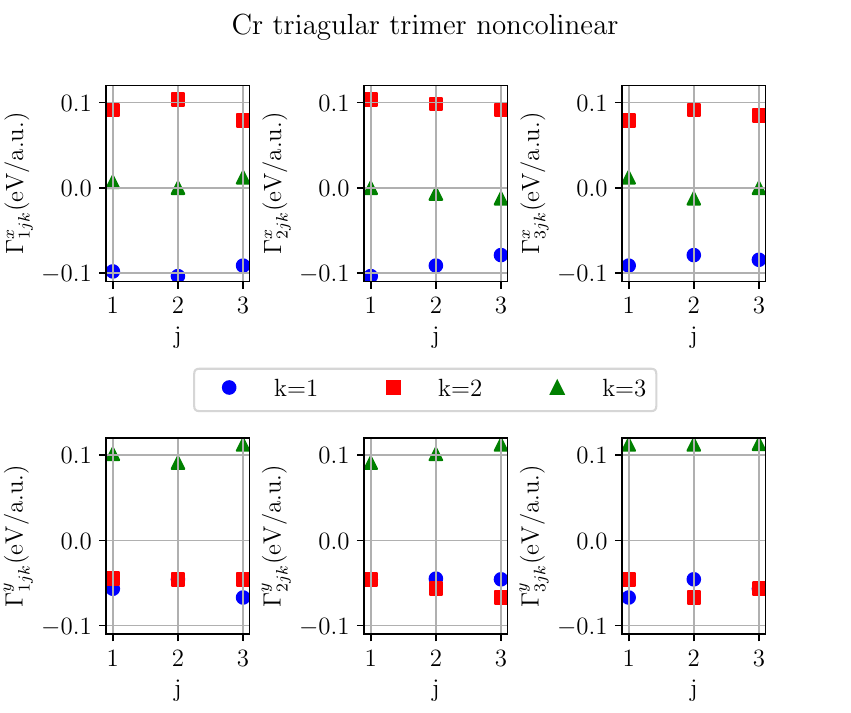}    
    \caption{The spin-lattice parameter $\Gamma_{ijk}^{\mu}$ for the triangular trimer of Fe (top panel) and Cr (middle and bottom panel). For the Cr, we present the spin-lattice parameter calculated from two different magnetic configuration: ferromagnetic (middle panel) and Néel state (bottom panel). }
    \label{fig:gamma}
\end{figure}

For Cr, we have $J_{12}=\unit[-143.9]{meV}$, force constants are $\Phi_{12}^{xx}=\unit[-0.81]{eV/a.u.^2}$,$\Phi_{12}^{xy}= \unit[0.13]{eV/a.u.^2}$, and $\Phi_{12}^{yy}=\unit[-0.07]{eV/a.u.^2}$. It is important to point out that representing these interaction in an irreducible form is only possible due to symmetries available in the ground state. It can be shown in this case that the pair interactions can be transformed into another. Breaking the symmetries in the magnetic and lattice ground state breaks also the symmetry relations of the pair couplings. The same holds also for the exchange striction terms, which are shown for two high-symmetry cases in Fig.~\ref{fig:gamma}. Important to notice are the points where $i=j$, which are finite for these nanostructures. We checked also nanostructures with periodic boundary conditions and obtained zero for parameters where $i=j$ (data not shown here). The magnetic configuration cases are the FM state and the N-AFM state. Remarkably, both magnitude and relation between different sites are significantly different. As already shown in Ref. \cite{streib_adiabatic_2022}, magnetic states far away from the actual ground state (here the Néel state) show stronger fluctuations when the internal degrees of freedom are varied. This causes larger values of $\Gamma$ in the ferromagnetic Cr trimer. Furthermore, the distance between the atoms turned out to be different for the FM state compared to the Néel state, after the lattice relaxation process, which in turn impact electron orbital overlaps that impacts the strength of the spin-spin coupling and variations of the spin-spin coupling. These strong changes of $\Gamma$ have an impact also on the dynamics, however, in semi-classical dynamics often only the values of the ground state are used. This aspect will be discussed further in more details in the next section where these parameters are used to perform spin-lattice dynamics and compared to our tight binding spin-lattice dynamics implementation.

\begin{figure}
    \centering
    \includegraphics[width=1\columnwidth]{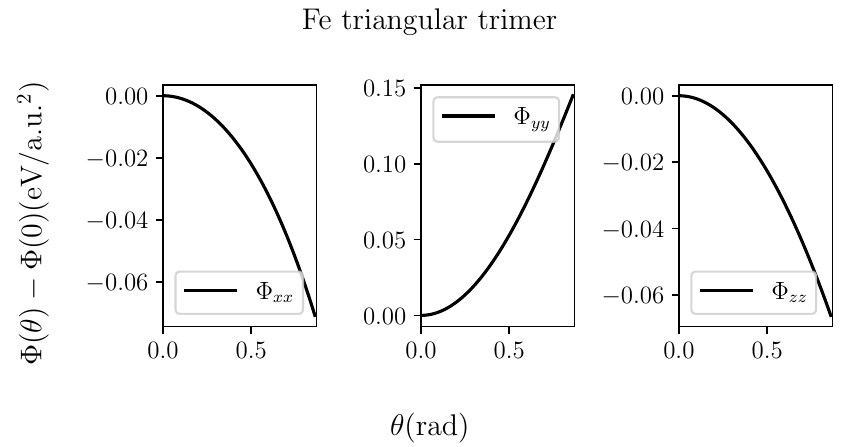}\\
    \includegraphics[width=1\columnwidth]{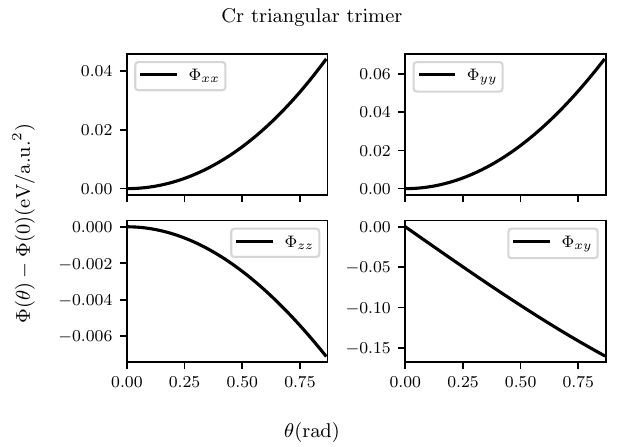}\\
    \caption{Onsite force constant matrix for the rotated atom in the Fe and Cr triangular trimer, top and bottom, respectively. }
    \label{fig:trimer}
\end{figure}

It has been heavily discussed in the literature how magnetic interactions are dependent on the underlying magnetic configuration of the system \cite{szilva_interatomic_2013, cardias_first-principles_2020,streib_exchange_2021,szilva2022quantitative}. The same question arises with respect to the force constants: do the force constants depend on the magnetic state? For instance, Ref.~\cite{PhysRevLett.113.165503} concluded that this dependence is neglegible. In order to investigate such dependence, we calculated the onsite force constants as a function of the magnetic configuration. The magnetic moment of a single atom, sitting along the y-axis, is rotated around the z-axis while the onsite force constant matrix is calculated (see Fig.~\ref{fig:schematic}). Given the nature of the onsite term, Eq.~(\ref{eq:sumrule}), the results can be interpreted as how the pairwise force constants of the other atoms react to the magnetic moment rotation of the atom considered. The rotation is done taking the ground state as reference, i.e. FM and N-AFM states for Fe and Cr, respectively. That magnetic dependence can be seen as the emergence of high-order terms in the magnetic moment, as the ones described in Eqs.~\eqref{eq:phi1}-\eqref{eq:phi2}. In the cases studied here, the only components that are not zero are $\Phi^{xx}$, $\Phi^{yy}$ and $\Phi^{zz}$. Particularly for Cr, given the noncollinear magnetic ground state, the $\Phi^{xy}=\Phi^{yx}$ is also not zero and present a linear dependence as opposed to the quadratic one seen in the other terms (Fig.~\ref{fig:trimer}). It is worth mentioning that components $\Phi^{zz}$ and $\Phi^{xy}=\Phi^{yx}$ (for the Cr case) are zero if the system is in a magnetic ground state. The changes are approximately $2\%$ for the $\Phi^{xx}$ and $1\%$ for the $\Phi^{yy}$ components for Fe. 
For Cr, the values are $8\%$ and $5\%$ for $\Phi^{xx}$ and $\Phi^{yy}$, respectively. \textbf{These percentages were obtained comparing the $\Phi$'s between $\theta=\unit[0]{rad}$ and $\theta=\unit[0.87]{rad}$}. The element dependence of these interactions is unclear since it depends on a variety of factors. Nevertheless, our examples show clearly that terms like \eqref{eq:phi1} and \eqref{eq:phi2} should be considered in semi-classical dynamics.


\subsection{Nanochain}
Here, we considered a nanochain with $10$ atoms along the x-axis, as shown in Fig.~\ref{fig:schematic}. Although being a simple one-dimensional example, there are a set of experimental data of nanochains supported on a class of metallic surfaces whose ground state is noncollinear~\cite{laszloffy_electronic_2021,laszloffy_magnetic_2019,xiao_formation_2006}.

\begin{figure}
    \centering
    \includegraphics[width=1\columnwidth]{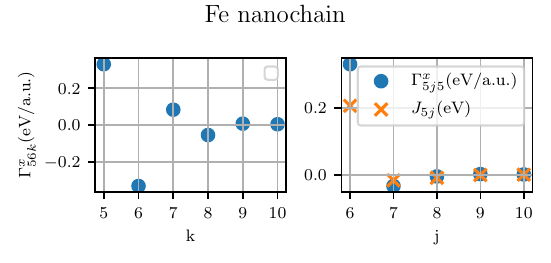}\\
    \includegraphics[width=1\columnwidth]{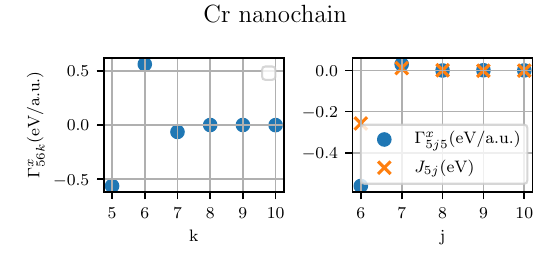}\\
    \caption{Spin-lattice parameters $\Gamma_{ijk}^{\mu}$ (exchange striction) and isotropic spin-spin exchange parameter $J_{ij}$ for the nanochains of Fe (top) and Cr (bottom). On the left panel, the atom 5 is taken as the reference atom while varying the $k$-th atom. On the right panel, the atom 5 is also taken as reference while the $j$-th atom is the one varying. Still on the right panel, for the spin-lattice parameter, the $k$-th atom is also being fixed.  }
    \label{fig:nanochain-par}
\end{figure}

On Fig.~\ref{fig:nanochain-par} we show the calculated exchange striction $\Gamma_{ijk}^{\mu}$ and the isotropic spin-spin exchange parameter $J_{ij}$ for the nanochain. In the nanochain case, only the x-component of the spin-lattice parameter, $\Gamma_{ijk}^{x}$, is not zero by symmetry. On the left side of the figure, we keep $i$ and $j$ fixed and vary $k$, which is the atom being displaced, from the $i-$th site ($\Gamma_{565}^{x}$) itself until the end of the chain, $k=10$. For that case, Fe presents a bigger sensitivity to lattice distortions far away from the local pair $\{i,j\}$, as oppose to the Cr nanochain case which the interaction shows itself more localized. On the right side, we keep $i$ and $k$ fixed to 5 and vary $j$ until $10$. For such case, the spin-lattice interaction for both Fe and Cr systems are localized with only relevant terms being the next-nearest and next next-nearest neighbors. For the isotropic spin-spin interaction, the relevant terms are also for the next-nearest and the next next-nearest neighbors. Regarding the force-constants $\Phi_{lk}^{\mu\nu}$, only the components $xx$, $yy$ and $zz$ are not zero in the nanochain case, with $yy$ and $zz$ having the same magnitude. The values are $\Phi_{55}^{xx}=\unit[4.72]{eV/a.u.^2}$ for the Fe and $\Phi_{55}^{xx}=\unit[3.39]{eV/a.u.^2}$ for Cr, while the other components are $\Phi_{55}^{yy}=\Phi_{55}^{zz}=\unit[-0.12]{eV/a.u.^2}$ and $\unit[-0.34]{eV/a.u.^2}$ for Fe and Cr, respectively. It should be noted that the force constants are calculated for the respective systems magnetic ground state. The magnetic dependence of the force constants having the ground state as reference is also verified for this case (data not shown).  

The above analysis of the interactions shows clearly that a precise description of the parametrized total energy \eqref{eq:adiabatic energy} by a defined Hamiltonian, e.g. Eq.~\eqref{eq:sldham}, is complex and can vary from material to material. Based solely on the values of $J$'s, $\phi$'s, and $\Gamma$'s, it is challenging to gauge the extent of the real-world impact that, e.g. the orientation of magnetic moments has on lattice dynamics or the lattice displacements have on the magnetic ground state. To delve deeper into assessing the consequences of these alterations, we conducted semi-classical spin-lattice dynamics analysis in Sec.~\ref{res:semisld}. This involved employing a classical spin-lattice Hamiltonian with fixed parameters and contrasting these outcomes with those obtained from a tight-binding spin-lattice approach. In the following section, we present the results of this comparative analysis.

\section{Spin-lattice dynamics of nanoclusters \label{sec:sldres}}
We conducted a comparative study by performing spin-lattice dynamics using two different approaches: the tight-binding (TB) model described above, and a spin-lattice Hamiltonian. In the former, we directly computed forces and the effective magnetic field from the electronic structure. In the latter, we computed the parameters of the spin-lattice Hamiltonian, assuming they remained constant during the dynamical process.

As discussed above, the semi-classical dynamics requests the energy parametrization from the ground state and, thus, misses effects that arbitrary configuration $\left\{\boldsymbol{e}\right\}$ and $\left\{\boldsymbol{u}\right\}$ are doing to the parameters, when not considering even larger expansions of the total energy in a parametrized Hamiltonian. The tight binding based dynamics, on the other hand, is able to calculate precisely fields and forces for these arbitrary configuration. Since the magnetic moment lengths are self-consistently determined at every time-step, the TB spin-lattice dynamics includes also longitudinal moment fluctuations, which are not addressed at all in typical semi-classical dynamics \cite{eriksson_atomistic_2017}. These fluctuations, however, will mainly influence the precession frequency of the moments, rather then the actual ground state.

\subsection{Dynamics via tight-binding model}

When focus extends beyond dynamics near an equilibrium state, as is often the case, utilizing tight-binding spin-lattice dynamics becomes important, at least for the systems considered here. This is especially pronounced when commencing from a randomly chosen initial state, a scenario frequently encountered when seeking to ascertain the precise ground state of a nanostructure. The TB spin-lattice dynamics calculations began with initially disordered magnetic configurations for both the triangular trimer and nanochain systems. Specifically, for the trimer, we investigated spin-lattice dynamics in various scenarios. These scenarios included comparing dynamics with and without fixing the magnetic moment orientation, which was initiated from a disordered magnetic configuration.

\begin{figure}[htp]
    \centering
    \includegraphics[width=1\columnwidth]{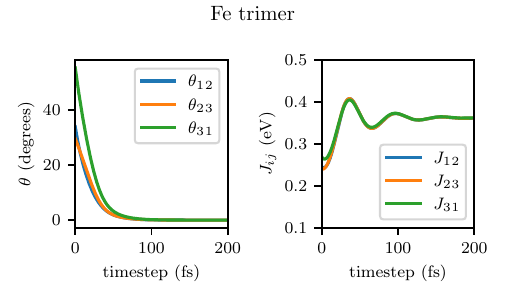}
    \includegraphics[width=1\columnwidth]{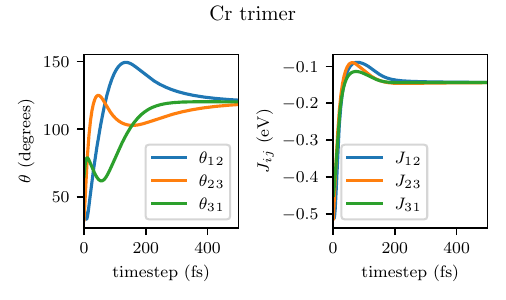}
    \includegraphics[width=1\columnwidth]{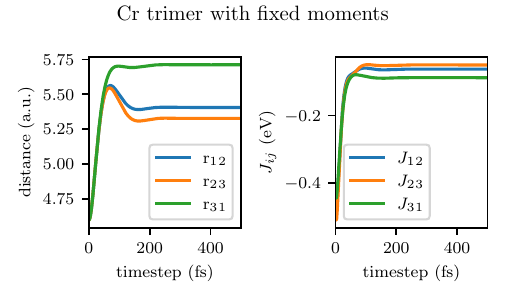}
    \caption{The angle between the magnetic moments of the Fe and Cr triangular trimer as a function of time in femtoseconds, as well as their respective exchange coupling parameters calculated at every time step. In the bottom panel, we show the results when the magnetic moments are considered fixed (lattice dynamics only), for the Cr triangular trimer case. The disordered magnetic moments induce a distortion in the lattice.}
    \label{fig:trimerjij}
\end{figure}

When the magnetic moments are fixed (lattice dynamics only performed), the results show that the magnetic disorder induces a distortion in the systems as the atomic sites relax to their lowest energy positions, as shown in Fig.~\ref{fig:trimerjij}, lower panel. This is due to the force induced by the non-collinearity. Such forces are, e.g., due to the exchange striction term in Eq.\eqref{eq:sldham}. However, forces calculated from the tight binding method do not fully agree to them (data not shown here) since they contain also higher order terms, such like forces coming from terms of the form \eqref{eq:phi1} and \eqref{eq:phi2}. This distortion effect is more dominant for Cr than for Fe (data not shown here) with a change in the distance up to $\unit[1]{a.u.}$ and has also impact on the magnetic exchange coupling, making it spatially anisotropic. 

\begin{figure*}[htp]
    \centering
    \includegraphics[width=1\columnwidth]{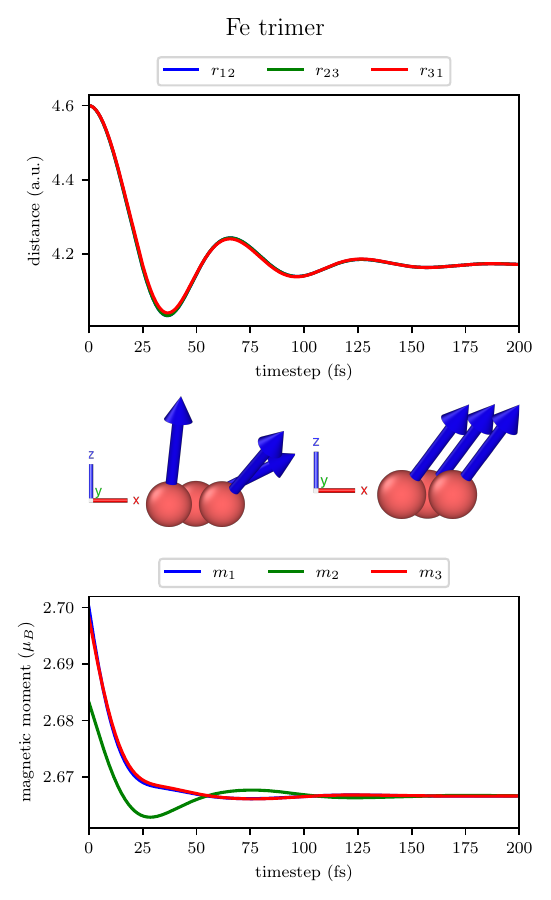}
    \includegraphics[width=1\columnwidth]{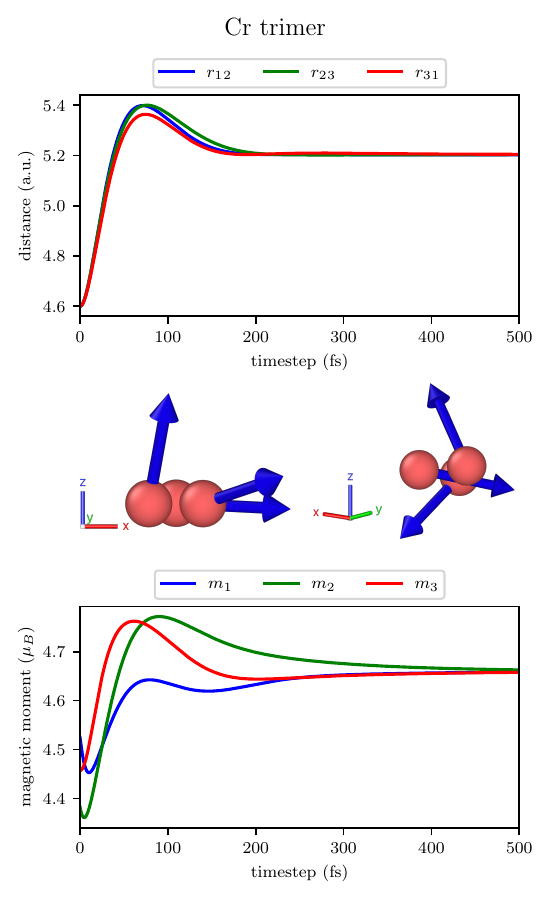}
    \caption{Spin-lattice dynamics for the triangular trimers of Fe and Cr. For these calculations, both the spin and lattice damping parameters are set to 0.05. On the left, we show the Cr triangular trimer relaxation process for the magnetic moment length (top) and the atomic positions (down). The figures in the middle denote the start configuration (left) and the final configuration (right). Note that the relaxation processes have different time scales, where the atomic positions relaxes at a faster rate than the magnetic moment lengths. The Fe triangular trimer is shown on the left, where magnetic moment lengths and atomic positions relaxes at a similar rate.}
    \label{fig:trimersld}
\end{figure*}

Now, if the magnetic moments are allowed to relax together with the atomic sites, i.e. the complete spin-lattice dynamics is performed, the systems (triangular trimers) find their true magnetic and molecular ground states, which in all cases correspond to the atomic sites sitting in the vertices of an equilateral triangle and the magnetic moments having a ferromagnetic ordering for Fe, while having an antiferromagnetic Néel state for the case of Cr (see the spin configurations in Fig.~\ref{fig:trimersld}). The trajectory of the spins towards this ground state is shown in Fig.~\ref{fig:trimerjij}, upper and center panel. Although we used the same energy dissipation parameter for both materials, the relaxation time of the magnetic moments is much larger for Cr compared to Fe and in the order of hundreds of femtoseconds. Similar isotropic spin-spin exchange constants $J_{ij}$ argue also for comparable timescales. It can be debated that during the relaxation process, the Cr magnetic moments cover a wider range of the configurational space, making higher order coupling mechanism to most likely play a more crucial role during the process for the Cr case as opposed to the Fe case. However, these couplings are hard to resolve for the constrained dynamics.



Additional insight into the distinct timescales observed can be obtained by examining both the magnetic moment lengths and lattice displacements, as depicted in Fig.~\ref{fig:trimersld}. For Cr, the magnetic moment length exhibits more significant variations, reaching up to $0.4\mu_B$, in contrast to Fe, where the variations are limited to $0.04~\mu_B$. Moreover, Cr inherently possesses larger magnetic moments, which in turn results in a lower precession magnetic field and, consequently, longer relaxation times. Analyzing lattice displacements reveals another disparity between Fe and Cr. In the case of Cr, the relaxation times of the lattice closely match those of the spins. This suggests a stronger spin-lattice coupling in Cr compared to Fe. Furthermore, around the $\unit[100]{fs}$ mark in the case of Cr, there's a notable divergence in the amplitudes of atomic distances, despite an initially symmetric atomic position. This discrepancy hints at an energy transfer between the magnetic moments and the lattice in Cr, likely due to high-order terms naturally considered in the TB approach and not taken into account via spin-lattice Hamiltonian, a phenomenon much weaker in the case of Fe.

\begin{figure*}[htp]
    \centering
    \includegraphics[width=1\columnwidth]{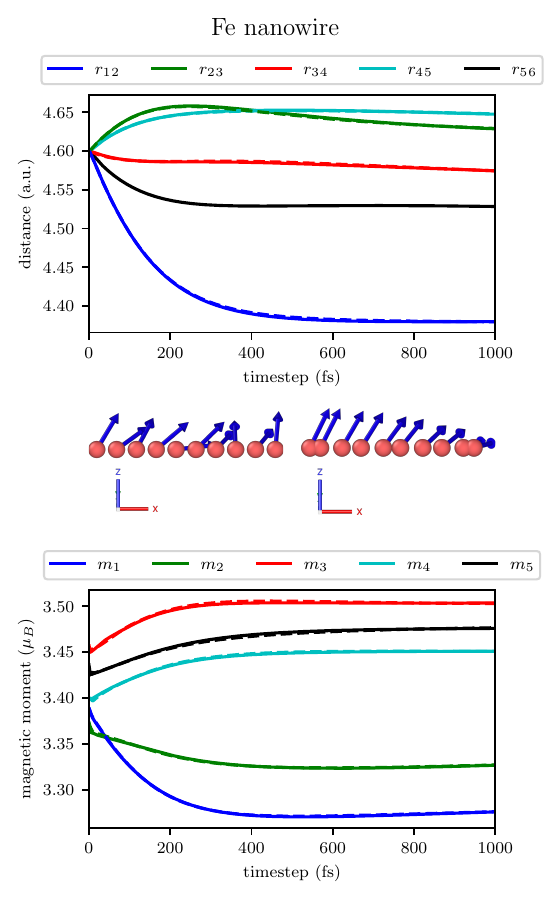}
    \includegraphics[width=1\columnwidth]{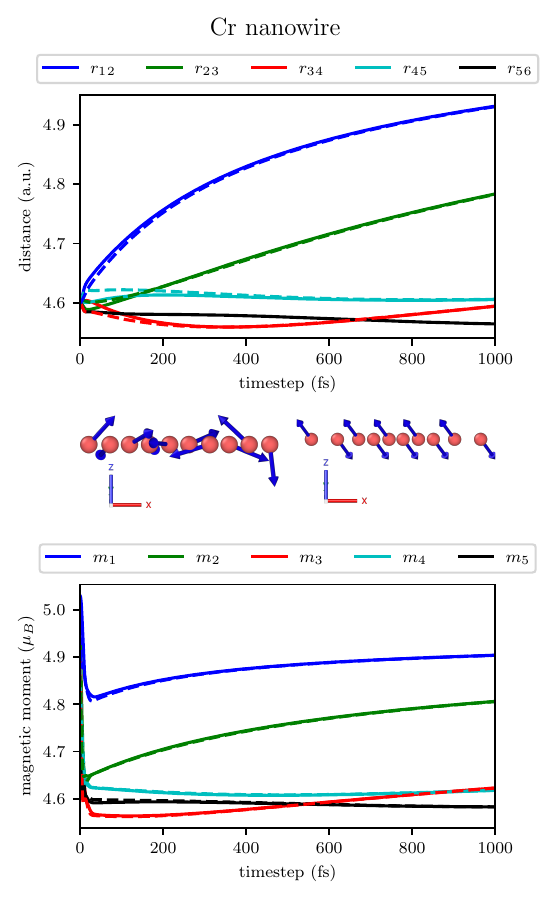}
    \caption{Spin-lattice dynamics for the nanochains with 10 atoms of Fe and Cr. For these calculations, both the spin and lattice damping parameters are set to 0.5. The figures in the middle denote the start configuration (left) and the final configuration (right). The full lines concern the atoms in the left half of the chain, while the dashed lines stand for the right half. One can note that during the first femtoseconds of simulation, they can be easily distinguished due to the disordered magnetic configuration.}
    \label{fig:nanochainsld}
\end{figure*}

The nanochain case is a slightly more complicated example compared to the trimer. This is due to the fact that now there are long range interactions that can compete with each other and may lead to a set of complex magnetic configurations, in case of spin-dynamics, and complex structural composition, in case of lattice-dynamics. We started the nanochains with a random magnetic configuration and the atoms sitting along the x-axis with a spacing of $\unit[4.6]{ a.u}$ between them, as can be seen in Fig.~\ref{fig:nanochainsld}. In the case of the nanochain with $10$ atoms, the first phenomenon which can be observed is the tendency of nearest neighbors atoms to occupy positions closer to each other. This is the so-called Peierls transition or Peierls distortion~\cite{little_possibility_1964}. Because electrons are free to move, it creates charge density waves which induce distortion to atomic positions. This has been observed before in weakly coupled molecular chains~\cite{thorne_chargedensitywave_1996}. As these atoms get closer to each other, their magnetic moments start to behave similarly, i.e. they act as a magnetic sub-lattice. For instance, for both Fe and Cr, this distortion leads to a dimerization of the atomic sites and their magnetic moment directions are then collinear between themselves, assuming a ferromagnetic alignment for the Fe and an antiferromagnetic alignment for the Cr case. Moreover, in the Fe case, these dimerized magnetic moments form a spin-spiral (with a long wave-length, as seen in Fig.~\ref{fig:nanochainsld}) as the magnetic ground state.  The phenomenon of dimerization was also discussed in Ref.~\cite{fransson_microscopic_2017}, were the authors highlight the presence of a dominant bilinear spin-lattice coupling term as the primary mechanism linking the spin and lattice behavior. Notably, this coupling term, which plays a crucial role in understanding the interplay between spin and lattice dynamics, is not considered in the semi-classical Hamiltonian \eqref{eq:sldham}. The omission of this term is due to the global frame defined in Ref.~\cite{streib_exchange_2021}, where its inclusion would disrupt the time reversal symmetry of the system. Consequently, the semi-classical description provided by \eqref{eq:sldham} would fail in capturing the dimerization effect. This underscores the distinct advantage of the here proposed tight-binding method, which accommodates the bilinear and other possible spin-lattice coupling term, enabling a more comprehensive analysis of the dimerization phenomenon. An infinitely long one-dimensional chain must transform under a Peierls distortion, but it is unclear if one-dimensional chains of limited size must distort as well. Most likely this is system dependent, but we note that the results presented here are consistent with what one would expect for infinite sized chains. 

Since the nanochain system has an even number of atoms, we can compare the left half with the right half. In Fig.~\ref{fig:nanochainsld}, we compare the different sides with the same color, but different linestyles. While the distance between pairs of atoms and their respective magnetization is shown with a full line for the left side, the right side is represented with a dashed line. Concerning the time evolution of the distances between different pairs of atoms, it is expected that both full and dashed lines, representing equivalent pairs of atoms from different sides, to behave equally. Instead, one can verify that full and dashed lines can be distinguished, specially in the beginning of the simulation. This is due to the fact that although both sides have equivalent spacing between the atoms, their magnetic moments are not equivalent. It suggests a transfer of angular momentum between the magnetic moments and the lattice which leads to a slightly different lattice dynamics on each side. That effect is, as it was for the trimer system, stronger on Cr than on Fe.


\subsection{Dynamics via spin-lattice Hamiltonian}\label{res:semisld}

It has been widely discussed in the literature the concept between local and global Hamiltonian \cite{streib_exchange_2021,szilva2022quantitative}. While the former relies on describing locally the energy of a given system, the latter uses a Hamiltonian with a complete set of parameters capable of describing the total energy of the system at any point in the phase space. Although convenient to use the global approach, the Hamiltonian cannot be known \textit{a priori}. A way of avoiding this issue is to extract both ionic forces and effective magnetic field directly from the electronic structure as shown in the previous Sections. 

\begin{figure*}[htp]
    \centering
    \includegraphics[width=1\columnwidth]{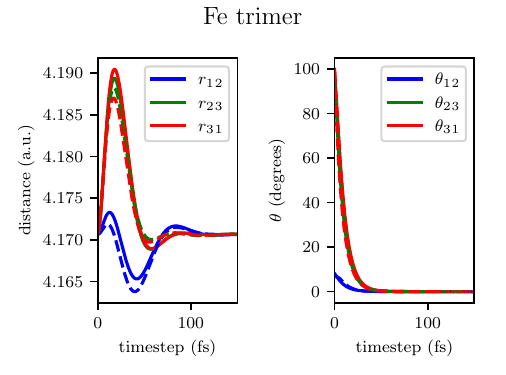}
    \includegraphics[width=1\columnwidth]{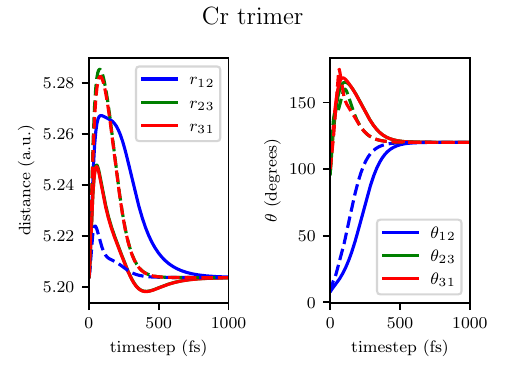}
    \caption{Comparison between the semi-classical (full line) and the tight-binding (dashed line) spin-lattice dynamics. The damping for both lattice and spin were set to 0.1. The parameters to the semi-classical spin-lattice Hamiltonian were calculated from the systems respective ground state, which are FM to the Fe trimer and Néel AFM for the Cr. Each system starts with disordered magnetic moments and in their respective ionic ground state.}
    \label{fig:semiclassxtb}
\end{figure*}

In this Section, instead of calculating the ionic forces and effective magnetic field directly from the electronic structure, we use the parameters described in Sec.~\ref{sec:resultssldham} as an input for a semi-classical spin-lattice Hamiltonian. From this Hamiltonian, the forces and effective field are calculated and used in the coupled spin-lattice dynamics. The goal is to determine validity of the semi-classical approach for a magnetically disordered system when comparing it to the tight binding spin-lattice dynamics. 
We started the simulation from the magnetic and structural ground state obtained from above described tight binding spin-lattice dynamics and by inducing an initial magnetic disorder while atoms sit in their relaxed atomic positions. If the system is in a magnetic ordering, since the ionic forces coming from the the exchange striction will be zero, the atoms will not move during the spin-lattice dynamics. However, the initial magnetic disorder gives rise to a net force and the atoms evolve. That can be seen in Fig.~\ref{fig:semiclassxtb}.

That movement can be understood as the angular momentum of the magnetic moments is transfer to the lattice, causing the atoms to move initially and going back to their ground state after some relaxation time. Surprisingly, we also observed such large discrepancies in the relaxation time between the Fe and Cr trimer as in the above discussed tight-binding dynamics. Since the semi-classical Hamiltonian includes the parametrization from the magnetic and atomic ground state, both the magnetization and lattice displacements relax toward this state. 

When comparing the tight-binding (dashed lines) with the semi-classical (solid lines) in Fig.~\ref{fig:semiclassxtb}, difference between both can be seen already in the early femtoseconds of the simulation and mostly in the displacements of the atoms.  This might be due to the emergence of high-order terms of the form \eqref{eq:phi1} and \eqref{eq:phi2} which we discussed in Sec.\ref{sec:resultssldham} but not explicitly included in the spin-lattice Hamiltonian \eqref{eq:sldham}. Surprisingly, the magnetic moments in the Fe trimer do not show much difference between the different methods. The discrepancy for the Cr trimer case is much more apparent. This is both in the relaxation time of the respective degree of freedom as well as in the magnitude of the perturbation. The lattice displacements are more significant in the tight binding case compared to the semi-classical model. This behavior can be comprehended by examining the exchange striction constants $\Gamma$, as demonstrated in Fig.~\ref{fig:gamma}. It is evident that $\Gamma$ exhibits an upward trend when transitioning from the Néel state to a ferromagnetic state. This increase in $\Gamma$ facilitates from the movement of spins that the atoms more readily displaced in comparison to the Néel state. An increase in the parameters also causes a growth in the precession frequency and, thus, relaxation times become smaller compare to the semi-classical case where the $\Gamma$'s of the Néel state are used. Furthermore, besides the pronounced variations in exchange striction, it is worth noting that other higher-order terms may also contribute to this disparity observed between tight-binding and semi-classical dynamics.


\section{Summary and Discussions \label{sec:sumdis}}


We have developed a fully integrated spin-lattice dynamics approach rooted directly in the tight-binding electronic structure. Unlike traditional methods that relies on a parametrized classical spin-lattice Hamiltonian, ours extracts molecular forces and the effective magnetic fields straight from the tight-binding structure via the Hellman-Feynman theorem and a self-consistent constraining field, respectively.

Our investigation into semi-classical Hamiltonians for Fe and Cr trimers and nanochains revealed significant variations in the Hamiltonian parameters based on the magnetic reference state. This aligns with literature findings about the magnetic texture influence on the magnetic interactions~\cite{streib_exchange_2021,jacobsson_efficient_2022,cardias_first-principles_2020,cardias_bethe-slater_2017,kvashnin_microscopic_2016}. More specifically, our tight-binding spin-lattice dynamics suggests the profound influence of magnetic texture on the Cr trimer case, whose ground state is non-collinear. In this case, parameters calculated from the ferromagnetic configuration as reference cannot correctly describe the system's dynamics. Still, the parameters calculated from the ground state describe the spin-lattice dynamics correctly only locally. i.e. for small variations of the magnetic moment orientation around the ground state. We argue that this scenario can be strongly present in sytsems whose magnetic ground state is non-collinear, such as 2D Kagomé magnets~\cite{2dmmrev}. Moreover, small differences can also be seen for the other systems.

The advantages of our present method is evident. It treats spin and lattice dynamics simultaneously, directly from the electronic structure, differing from previous methods like those in Ref.~\cite{stockem_anomalous_2018}. The self-consistent electronic structure calculation at each step enables consideration of magnetic moment length relaxation, although certain abrupt changes can only be approximated at this stage.

Looking ahead, we are focusing on incorporating spin-orbit coupling effects, anticipating even more pronounced variances between tight-binding and semi-classical dynamics, especially with non-collinear effects at finite temperatures. In essence, our new approach offers a complete tool for probing spin-lattice dynamics from first principles across diverse materials.

\section{Acknowledgments}
This work was financially supported by the Knut and Alice Wallenberg (KAW) Foundation through Grant Nos.\,2018.0060, 2021.024, and 2022.0108.
O.E.~acknowledges support by the Swedish Research Council (VR), the European Research Council (854843-FASTCORR), eSSENCE and STandUP. 
D.T.~acknowledges financial support from the Swedish Research Council (Vetenskapsrådet, VR), Grant No. 2019-03666.
A.D.~acknowledges financial support from the Swedish Research Council (Vetenskapsrådet, VR), Grant Nos. 2016-05980 and 2019-05304.
A.D. and O.E. acknowledge support from 
the Wallenberg Initiative Materials Science
for Sustainability (WISE) funded by the Knut and Alice
Wallenberg Foundation (KAW),
R.C acknowledges financial support from FAPERJ - Fundação Carlos Chagas Filho de Amparo à Pesquisa do Estado do Rio de Janeiro, grant number E-26/205.956/2022 and 205.957/2022 (282056). 
The computations were enabled by resources provided by the National Academic Infrastructure for Supercomputing in Sweden (NAISS) and the Swedish National Infrastructure for Computing (SNIC) at NSC and PDC, partially funded by the Swedish Research Council through grant agreements no. 2022-06725 and no. 2018-05973.


\appendix 
\section{Tight-binding model\label{app:tbmodel}}
To obtain the electronic structure, we use a tight-binding model \cite{Grotendorst09} based on the Slater–Koster
parameterization \cite{PhysRev.94.1498} and solve

\begin{equation}
    \mathcal{H}\mathbf{C}_n = \varepsilon_n \mathcal{S}\mathbf{C}_n\label{eq:tb}
\end{equation}
$\mathbf{H}$ and $\mathbf{S}$ are the real-space Hamiltonian and overlap matrices, respectively, while $\mathbf{C}_n$ is the eigenstate $n$ with energy $\varepsilon_n$. The states are represented in a spd basis set. Due to the presence of the definite and positive $\mathbf{S}$ matrix, Eq.~\eqref{eq:tb} is called a generalized eigenvalue problem. To transform it to an orthogonal problem, we apply Cholesky decomposition method. 

Hamiltonian consists of a hopping term $\mathcal{H}^0$, a local charge neutrality term $\mathcal{H}^{lcn}$, and a Stoner term $\mathbf{H}^{st}$. The Slater-Koster parameter enter into the hopping term and are furthermore represented via a polynomial expansion in the distance between the atoms $\mathbf{r}_{ij}$, multiplied with a Slater-type orbital. This expansion was motivated by Mehl and Papaconstantopoulos \cite{mehl_applications_1996} as well as further refined and applied in Ref. \cite{barreteau_efficient_2016}. It has been demonstrated that the utilization of a tight-binding model offers a computationally efficient and accurate portrayal of transition metal elements and alloys \cite{barreteau_efficient_2016}. This model is effective in describing both collinear and noncollinear magnetic arrangements \cite{cardias_spin_2021} as well as lattice dynamics and phonon \cite{Papaconstantopoulos_2003}, and its validity has been established by comparing its results with those obtained from DFT calculations. The parameter for the hopping part of the Hamiltonian are obtained from Ref. \cite{Papaconstantopoulos_2014}.

The local charge neutrality and the Stoner term are defined according to Ref. \cite{barreteau_efficient_2016} and \cite{streib_adiabatic_2022}, respectively. It is important to mention that the Mulliken transformation \cite{Schena2010} need to be applied to both terms. For the local charge neutrality constant we use $U_{lcn} = \unit[5]{eV}$. The parameters for the Stoner Hamiltonian are obtained from Ref.\cite{Rossen2019} and we used $I_s=I_p=\nicefrac{I_d}{10}$ for the orbital resolved Stoner parameters. We add furthermore the constrain hamiltonian, as mentioned in the main text. 

Eq. \ref{eq:tb} is solved by exact diagonalization and self-consistently in the charge $n_i$, magnetic moment length $m_i$, and constraint field $\mathbf{B}^{const}_i$ using Anderssen mixing scheme and a self-consistent threshold error of $1\cdot 10^{-10}$. This implies up to $50$ iterations per self-consistent run. The dimension of the problem is $N=2N_{orb}N_{atom}$ with $N_{orb}$ being number of orbitals (here $N_{orb}=18$) and $N_{atom}$ being the number of atoms. 

\section{Tight-binding spin-lattice dynamics process\label{app:sld}}

In order to solve the coupled dynamics for magnetic moment \eqref{eq:llg} and lattice displacement \eqref{eq:md}, we need the effective field as well as the lattice force at each site $i$. Following Ref.\cite{streib_equation_2020}, the effective field is $\mathbf{B}_{i}^{\text{eff}}=-\mathbf{B}_{i}^{\text{con}}$, the opposite of the constraint field. The lattice forces are calculated via the Hellman-Feynman theorem (see Appendix \ref{sec:forces}). We carefully checked the obtained effective fields and forces with numerical finite differences of the total energy. It is important that the total energy needs to be corrected by double counting terms coming from the Stoner and local charge neutrality term.

The dynamics equations \eqref{eq:llg} and \eqref{eq:md} are solved via implicit midpoint method, described in Ref.~\cite{hellsvik_general_2019}. The implicit solver uses about $5$ iteration step to be converged within the threshold of $1\cdot10^{-10}$. Nevertheless, each step requires a self-consistent solution of the electronic system in order to obtain the obtained the constraint field and the forces for the updated magnetic and lattice configuration. This makes the entire procedure computational demanding and limits the method so far to systems with maximum 100 atoms. 

It should be noticed that the proposed method fails when abrupt large changes in the self-consistent variables occur. This would generate extra terms in the electron Hamiltonian proportional, e.g. to $\nicefrac{\partial m_i}{\partial u_k^\alpha}$ and others, as proposed for the spin case in Ref.\cite{streib_equation_2020}.

\section{Hellmann-Feynman theorem and tight-binding moecular forces}
\label{sec:forces}
Let us consider a basis written in terms of linear combination of atomic orbitals (LCAO) 

\begin{equation}
    \ket{\Psi} = \sum_{i\ell}{c_{i\ell}\psi_{i\ell}},
\end{equation}
with $i$ as the atom index and $\ell$ the orbital type. Then one can write the overlap matrices as

\begin{equation}
    \mathcal{S}_{i\ell, j\ell'} = {\langle{\psi_{i\ell}}|{\psi_{j\ell'}}\rangle},
\end{equation}
with the secular equation being

\begin{equation}
  \mathcal{H}_{ij}c_{j}=\epsilon\mathcal{S}_{ij}c_{j}, 
  \label{eq:secular}
\end{equation}
where we omit from this point on the orbital indexes $\ell$ and $\ell'$ for simplicity. Here, $\mathcal{H}_{ij} = \bra{\psi_{i}}\mathcal{H}\ket{\psi_{j}}$. Thus, the normalization condition can be written as

\begin{equation}
    {\langle{\Psi}|{\Psi}\rangle} = c_{i}^{*}\mathcal{S}_{ij}c_{j} = 1,
    \label{eq:norm1}
\end{equation}
where a sum is implicit in the repeated indexes. Similarly, one can write

\begin{equation}
    \bra{\Psi}\mathcal{H}\ket{\Psi} = c_{i}^{*}\mathcal{H}_{ij}c_{j} = \epsilon.
    \label{eq:norm2}
\end{equation}

From Eqs.~(\ref{eq:norm1}) and (\ref{eq:norm2}), one can derive the following relations

\begin{align}
    \frac{\partial  {\langle{\Psi}|{\Psi}\rangle}}{\partial \lambda} = \frac{\partial}{\partial \lambda}(c^{*}_{i}\mathcal{S}_{ij}c_{j}) &= 0 \nonumber \\
    \frac{\partial c^{*}_{i}}{\partial \lambda}\mathcal{S}_{ij}c_{j} + c_{i}^{*}\mathcal{S}_{ij}\frac{\partial c _{j}}{\partial \lambda} &= -c_{i}^{*}\frac{\partial\mathcal{S}_{ij}}{\partial \lambda}c_{j}
    \label{eq:rel1} 
\end{align}

and
\begin{align}
    \frac{\partial \epsilon}{\partial\lambda} = c^{*}_{i}\frac{\mathcal{H}_{ij}}{\partial \lambda}c_{j} + \epsilon\left[\frac{\partial c^{*}_{i}}{\partial\lambda}\mathcal{S}_{ij}c_{j}+ c^{*}_{i}\mathcal{S}_{ij}\frac{c_{j}}{\partial\lambda}\right].
    \label{eq:rel2}
\end{align}

According to the Eqs.~(\ref{eq:secular}) and (\ref{eq:rel2}), we can finally write

\begin{align}
   \vec{F}_{k} = -2\left[\sum_{\ell}c_{k\ell}\sum_{j\neq k}\sum_{\ell'}c_{j\ell'}\left(\frac{\partial \mathcal{H}_{k\ell,j\ell'}}{\partial \vec{R}_{k}}-\epsilon\frac{\partial \mathcal{S}_{k\ell,j\ell'}}{\partial \vec{R}_{k}}\right)\right],
\end{align}
where $\ell, \ell'$ are the orbital indexes and $\{k,j\}$ are the site indexes. Given the fact that both the Hamiltonian $\mathcal{H}_{k\ell,j\ell'}$ and the overlap matrix $\mathcal{S}_{k\ell,j\ell'}$ are a function of the Slater-Koster table in the NRL tight-binding approach, the derivative with respect to the atomic positions $\vec{R}_{k}$ have an analytical expression which is essentially the derivative of the Slater-Koster table with respect to the direction cosines. This was done in Ref.~\cite{jdziedzic2007}. More details on the tight-binding model can be seen fully in Refs.~\cite{barreteau_efficient_2016,Schena2010,cardias_spin_2021,streib_adiabatic_2022}.

\bibliography{bibfile/sldbib}

\end{document}